\documentclass[11pt,a4paper]{article}
\usepackage[margin=25mm]{geometry}
\usepackage{amsmath,amssymb,graphicx,booktabs,array,longtable}
\usepackage[numbers,sort&compress]{natbib}
\usepackage[expansion=false]{microtype}
\usepackage{caption}
\usepackage{placeins}
\usepackage[colorlinks=true,linkcolor=blue,citecolor=blue,urlcolor=blue]{hyperref}
\newcommand{\GeV}{\,\mathrm{GeV}}
\newcommand{\TeV}{\,\mathrm{TeV}}
\newcommand{\sigmav}{\langle\sigma v\rangle}
\newcommand{\unitrate}{\mathrm{cm^3\,s^{-1}}}
\newcommand{\mchi}{m_{\chi}}
\newcommand{\om}{\Omega_\chi h^2}
\newcommand{\MG}{M_{\rm GUT}}
\newcommand{\FO}{\mathrm{FO}}
\title{Neutralino Dark Matter in SU(5)\\ after H.E.S.S. and Direct-Detection Searches}
\author{M. Adeel Ajaib and Fariha Nasir\\[4pt]
\small Pennsylvania State University, Abington College,\\
\small Abington, PA 19001, USA}
\date{}
\begin{document}
\maketitle
\begin{abstract}
We examine the impact of gamma-ray line and direct-detection searches on neutralino dark matter in supersymmetric SU(5) with non-universal gaugino masses. Under the assumption that neutralinos constitute all halo dark matter, a nearly pure wino benchmark at $1\TeV$ is essentially excluded by the H.E.S.S. line limits for both cusped and cored density profiles. Higgsino and wino--Higgsino benchmarks are independently excluded by direct detection. In contrast, a bino benchmark near thermal saturation combines suppressed present-day line emission with scattering below the displayed experimental limits. Mixed bino--wino solutions illustrate the complementarity of the two searches and the dependence of the line interpretation on the Galactic halo profile. We relate these results to the GUT-scale gaugino hierarchy and the SU(5) scalar and Higgs boundary conditions. Thermally underabundant solutions require additional production to realize the full-density assumption. Comparing the calculated signals with published electroweakino predictions, we identify the spectrum-dependent corrections and collider tests needed to determine the surviving parameter space.
\end{abstract}

\section{Introduction}
Supersymmetric grand unified theories provide a framework in which the weak-scale sparticle spectrum and dark-matter observables can be related to a relatively small set of high-scale parameters. With conserved $R$ parity, the lightest neutralino is a dark-matter candidate when it is the lightest supersymmetric particle. Its phenomenology depends on the relative bino, wino and Higgsino contributions to its wave function. These contributions determine the available annihilation channels, the relevance of nearby sparticles at freeze-out, and the size of elastic-scattering and indirect-detection signals.

Independent $M_1$, $M_2$ and $M_3$ at $\MG$ permit light electroweakinos while the colored states remain heavy. Previous studies examined the SU(5) spectrum, relic abundance, direct detection and the muon magnetic moment~\cite{Ajaib2017SU5,Ajaib2017Status}. Related work demonstrated the role of light sleptons and non-universal gaugino masses~\cite{Ajaib2015GUT,Ajaib2014Split} and the collider signatures of different neutralino compositions~\cite{Ajaib2015Neutralinos}. Recent analyses of non-universal soft terms combine the Higgs mass, muon magnetic moment and dark-matter constraints~\cite{Ellis2024NonUniversal}; an SU(5) realization with a comparatively heavy gluino also examines direct-detection and collider prospects~\cite{Dong2025NUGM}. Our emphasis is the relation of the SU(5) scalar and Higgs boundary conditions to the composition-dependent scattering and line signals, without imposing a muon-$g-2$ interval on the main sample. A separate comparison highlights solutions in the interval $0\leq\Delta a_\mu\leq6.6\times10^{-10}$. We examine how the H.E.S.S. Inner Galaxy Survey and recent xenon direct-detection searches test the composition classes in this high-scale framework under a full-density interpretation. The resulting comparison identifies candidate spectra for further tests rather than a fully delimited region allowed by all collider and dark-matter searches.

The channels $\chi\chi\rightarrow\gamma\gamma$ and $\chi\chi\rightarrow Z\gamma$ produce narrow spectral features. The H.E.S.S. Inner Galaxy Survey provides observed limits in the TeV region and compares them with precision electroweak-multiplet predictions~\cite{HESS2026Lines}. Earlier H.E.S.S., MAGIC and Fermi-LAT line searches establish the broader experimental context~\cite{HESS2018Lines,MAGIC2023Lines,Fermi2015Lines}. Sommerfeld enhancement and radiative corrections are central to this interpretation~\cite{Hisano2004,Cohen2013,Fan2013,Baumgart2019,Rinchiuso2018,Beneke2020Higgsino,Beneke2022Spectrum}. Direct detection supplies an independent test, particularly when electroweak mixing enhances the Higgs coupling.

We assume that neutralinos supply the full halo density and use their calculated scattering and annihilation cross sections directly. The thermal relic abundance remains a separate diagnostic. Additional production can populate thermally underabundant neutralinos; nonthermal winos provide a well-studied example~\cite{Moroi2000,Baer2018FullDensity}. We distinguish these solutions from those approximately saturating the observed abundance through the calculated freeze-out process.

Our analysis connects SU(5) boundary conditions to neutralino composition, thermal abundance and detection observables. We compare fixed-order two-body rates with continuous experimental limits and with the published canonical electroweakino predictions. The distinction between these predictions matters most for compressed wino and Higgsino spectra. Their thermal abundances are also sensitive to electroweak nonperturbative effects~\cite{Beneke2016WinoRelic,Bottaro2022}.

The remainder of the paper is organized as follows. Section~\ref{sec:model} describes the model, numerical sample and selections. Section~\ref{sec:observables} defines the composition and detection observables. We present the results in Section~\ref{sec:results}, including benchmark points. Section~\ref{sec:conclusion} contains our conclusions.

\section{SU(5) Model and Scanning Procedure}\label{sec:model}
We consider the conventional SU(5) matter assignment, in which $Q$, $U^c$ and $E^c$ belong to the ${\bf 10}$ and $D^c$ and $L$ belong to the $\overline{\bf 5}$. The soft scalar boundary conditions are
\begin{align}
 m_Q=m_{U^c}=m_{E^c}&=m_{10}, &
 m_{D^c}=m_L&=m_{\overline 5}, \label{eq:scalar}\\
 m_{H_u}=m_{H_d}&=m_{\overline 5}, & A_0&=a_0m_{\overline 5},\label{eq:higgs}
\end{align}
at $\MG$, with the matter relations applied to all three families. The quantities in Eqs.~(\ref{eq:scalar})--(\ref{eq:higgs}) denote soft mass parameters, with the corresponding squared masses entering the scalar potential. The trilinear parameter is universal after factoring out the Yukawa couplings. {Table~\ref{tab:domain} specifies the parameter ranges employed in our analysis}. The scanning procedure and assumptions follow the SU(5) model discussed in Ref.~\cite{Ajaib2017SU5}.

The three gaugino inputs are independent and have signs
\begin{equation}
 M_1>0,\qquad M_2>0,\qquad M_3<0,\qquad \mu>0.\label{eq:signs}
\end{equation}
Non-universal gaugino masses can arise from supersymmetry-breaking $F$-terms that transform as non-singlets under the unified gauge group. Contributions from fields in different representations, or from more than one source of supersymmetry breaking, allow a variety of gaugino mass patterns at $\MG$. Motivated by these possibilities, we treat $M_1$, $M_2$ and $M_3$ as independent parameters, following the SU(5) framework of Ref.~\cite{Ajaib2017SU5}. The top-quark pole-mass input is $173.3\GeV$.

\begin{table}[tb]
\centering
\begin{tabular}{lll}
\toprule
Parameter & Nominal scan domain or fixed value \\
\midrule
$m_{10},\ m_{\overline5}$ & $0 \rightarrow 30\TeV$ \\
$M_1,\ M_2$ & $0 \rightarrow 5\TeV$ \\
$M_3$ & $-5 \rightarrow 0\TeV$ \\
$a_0=A_0/m_{\overline5}$ & $-3 \rightarrow  3$ \\
$\tan\beta$ & $2 \rightarrow 60$ \\
$m_{H_u},\ m_{H_d}$ & $m_{\overline5}$ \\
$\operatorname{sign}(\mu)$ & $+1$ \\
$m_t^{\rm pole}$ & $173.3\GeV$ \\
\bottomrule
\end{tabular}
\caption{GUT-scale scan ranges and other fixed inputs used in the analysis.}\label{tab:domain}
\end{table}

The spectrum is calculated with ISAJET/ISASUGRA~\cite{ISAJET} and passed in SLHA format to micrOMEGAs for relic-density and detection observables~\cite{micromegasIndirect}. The $\gamma\gamma$ and $Z\gamma$ channels use the one-loop \texttt{loopGamma} calculation~\cite{Boudjema2005Lines,micromegasManual52}.

We first apply the requirements in Table~\ref{tab:cuts}, including the Higgs mass, flavor observables, charged and colored sparticle thresholds, and neutralino mass ordering. These cuts define the baseline sample. Relic abundance and detection constraints are then considered separately, allowing their effect on each composition class to be identified. In particular, the imposed $103.5\GeV$ threshold is a chargino selection, not a lower bound of the same value on a bino LSP. A bino near $62\GeV$ cannot be rejected using that threshold alone. Prompt electroweakino searches depend on production rates, mass gaps and branching fractions; their simplified-model exclusions do not define a universal neutralino mass band. The ATLAS pMSSM interpretation demonstrates the importance of testing these ingredients together~\cite{ATLAS2024pMSSM}.

\begin{table}[tb]
\centering
\begin{tabular}{ll}
\toprule
Observable & Baseline requirement \\
\midrule
$m_h$ & $123\leq m_h\leq127\GeV$ \\
$\mathrm{BR}(b\rightarrow s\gamma)$ & $2.90\times10^{-4}\leq\mathrm{BR}\leq3.87\times10^{-4}$ \\
$\mathrm{BR}(B_s\rightarrow\mu^+\mu^-)$ & $1.95\times10^{-9}\leq\mathrm{BR}\leq3.43\times10^{-9}$ \\
$\mathrm{BR}(B_u\rightarrow\tau\nu)_{\rm MSSM}/\mathrm{BR}_{\rm SM}$ & $0.15$--$2.41$ \\
$m_{\widetilde g}$ & $\geq2200\GeV$ \\
$m_{\widetilde t_1}$ & $\geq700\GeV$ \\
All first- and second-family squark masses & $\geq2000\GeV$ \\
$m_{\chi_1^\pm}$ & $\geq103.5\GeV$ \\
$m_{\widetilde\tau_1}$ & $\geq93.2\GeV$ \\
Checked sparticle ordering & Neutralino is lightest \\
\bottomrule
\end{tabular}
\caption{Selections used to construct the baseline sample. Relic-density selections and direct-detection comparisons are discussed in the results section. The muon-$g-2$ interval is used only for the additional comparison in Fig.~\ref{fig:gm2subset}. The mass selections provide a common numerical baseline and do not establish compatibility with every collider search.}\label{tab:cuts}
\end{table}

\subsection*{Relic-Density Constraint}
For statements involving thermal abundance we additionally require a finite positive $\om$, a zero relic-calculation error flag, and neutralino identification by micrOMEGAs. We also verify finite nonnegative line rates and normalized neutralino fractions. These checks define the quality sample and remove entries with missing relic results, nonzero relic-calculation flags or a charged dark-matter candidate. Removing the chargino-labelled entries is necessary even though they pass the stored mass-ordering test; a rounded or numerically degenerate spectrum should not supersede the identity returned by the dark-matter calculation.

We consider two further subsets,
\begin{align}
\mathcal{S}_{\rm upper}:&\quad 0<\om\leq0.13,\label{eq:upper}\\
\mathcal{S}_{\rm sat}:&\quad 0.11\leq\om\leq0.13.\label{eq:saturation}
\end{align}
The upper-bound sample retains thermally underabundant solutions together with solutions near thermal saturation. For its underabundant members, the full-density detection interpretation is conditional on additional neutralino production. The narrower saturation sample identifies solutions whose calculated thermal abundance is approximately sufficient. These selections are nested: $\mathcal{S}_{\rm sat}\subset\mathcal{S}_{\rm upper}$. The relic selections characterize the freeze-out result; they do not set the normalization of the detection cross sections.

The reference abundance is $\Omega_{\rm DM}h^2=0.12$, motivated by the Planck result~\cite{Planck2018}. The bounds $0.11$ and $0.13$ specify the tolerance adopted in this analysis; they do not define the Planck experimental confidence interval. This tolerance permits a small excursion above the reference value. Solutions above the adopted upper ceiling remain outside the signal sample; no dilution mechanism is invoked to rescue them.

\section{Neutralino composition and detection observables}\label{sec:observables}
\subsection{Composition classes and mass diagnostics}
In the gauge-eigenstate basis we write
\begin{equation}
 \chi\equiv\widetilde\chi_1^0=N_{11}\widetilde B+N_{12}\widetilde W^0
 +N_{13}\widetilde H_d^0+N_{14}\widetilde H_u^0.\label{eq:mix}
\end{equation}
We define
\begin{equation}
 f_B=|N_{11}|^2,\qquad f_W=|N_{12}|^2,\qquad
 f_H=|N_{13}|^2+|N_{14}|^2,\qquad f_B+f_W+f_H=1.\label{eq:fractions}
\end{equation}
A point is bino, wino or Higgsino dominated if the corresponding fraction is at least 0.8. Otherwise, its two largest fractions determine the mixed label. Thus the composition labels are reproducible classifications, not assertions of exact gauge eigenstates. In particular, a mixed label does not require equal fractions.

The low-scale mass parameters $M_1(Q)$, $M_2(Q)$ and $\mu(Q)$ govern the neutralino matrix, whereas the independent input quantities are defined at $\MG$. We retain this scale distinction throughout. The resulting mapping is affected by renormalization-group evolution and electroweak symmetry breaking, so a single high-scale gaugino ratio need not determine the composition by itself. We use the joint ratio plane to display the regions actually sampled.

To characterize compression we define
\begin{equation}
 \Delta m_{\chi^\pm}=m_{\chi_1^\pm}-\mchi,\qquad
 d_i=\frac{m_i-\mchi}{\mchi}.\label{eq:compression}
\end{equation}
Small positive $d_i$ indicates a kinematic opportunity for coannihilation. It is not, on its own, a calculation of a process contribution to the effective freeze-out cross section. The same distinction applies to the resonance diagnostic $|m_A-2\mchi|/(2\mchi)$. We therefore describe compressed or near-resonant spectra without assigning a unique relic mechanism from masses alone.

\subsection{Full-density assumption and elastic scattering}
For the detection interpretation of points in $\mathcal{S}_{\rm upper}$ we assume
\begin{equation}
 \rho_\chi(\mathbf r)=\rho_{\rm DM}(\mathbf r).\label{eq:density}
\end{equation}
We consequently use the calculated spin-independent and spin-dependent neutralino--nucleon cross sections directly. The full-density assumption applies to both local scattering and the Galactic annihilation signal. For thermally underabundant points it requires additional neutralino production, such as a nonthermal contribution; its existence and cosmological consistency are assumptions outside the present calculation~\cite{Baer2018FullDensity}. The reported thermal abundance is not replaced by the observed abundance in the data tables. Recent calculations of Higgsino production from late-decaying scalar fields illustrate how a thermally underabundant electroweakino can attain the observed abundance, including coannihilation and Sommerfeld effects~\cite{Fukuda2025Nonthermal}. A corresponding production history has not been constructed for our spectra.

We compare the proton spin-independent cross section with the observed limits from LZ, XENONnT and PandaX-4T~\cite{LZ2025,XENONnT2025,PandaX2025}. The reference curves assume elastic, approximately isospin-conserving scattering and each experiment's standard halo model. The proton and neutron amplitudes of the Higgsino and wino--Higgsino benchmarks differ by less than 1\%, making this comparison appropriate for those points.

\subsection{Two-body gamma-ray lines and the H.E.S.S. convention}
For neutralinos annihilating at rest, the line energies are
\begin{equation}
 E_{\gamma\gamma}=\mchi,\qquad
 E_{Z\gamma}=\mchi-\frac{m_Z^2}{4\mchi}.\label{eq:energies}
\end{equation}
The on-shell $Z\gamma$ channel is available only for $\mchi\geq m_Z/2$. Below threshold we set its contribution to zero and do not interpret a negative value obtained by extending Eq.~(\ref{eq:energies}) as a photon energy. In the sample the below-threshold entries already have zero $Z\gamma$ rates.

For a self-conjugate particle and an integrated angular region, the two-body photon flux may be written
\begin{equation}
 \frac{d\Phi_\gamma}{dE}=\frac{J}{8\pi\mchi^2}
 \left[2\sigmav_{\gamma\gamma}\delta(E-\mchi)
 +\sigmav_{Z\gamma}\delta(E-E_{Z\gamma})\right],\label{eq:flux}
\end{equation}
where $J$ is the line-of-sight integral of the squared total dark-matter density over that region, identified with the neutralino density through Eq.~(\ref{eq:density}). The two photons in the $\gamma\gamma$ final state produce the factor of two. The photon-weighted quantity is $L_\gamma=2\sigmav_{\gamma\gamma}+\sigmav_{Z\gamma}$. The equivalent two-photon line convention used in the H.E.S.S. comparison is instead~\cite{HESS2026Lines}
\begin{equation}
 \sigmav_{\rm line}^{\FO}=\sigmav_{\gamma\gamma}^{\FO}
 +\frac{1}{2}\sigmav_{Z\gamma}^{\FO}=\frac{L_\gamma^{\FO}}{2}.\label{eq:line}
\end{equation}
All line figures and benchmark tables use Eq.~(\ref{eq:line}). The superscript FO identifies the fixed-order two-body prediction. No thermal-abundance multiplier is applied.

For the H.E.S.S. comparison we require $\mchi\geq300\GeV$. The fractional separation of the two lines is at most 2.3\% in this region. Figure~\ref{fig:line} displays the observed limits for Einasto, NFW, cNFW, FIRE-2 and Auriga density profiles, with the profile-dependent normalizations of Ref.~\cite{HESS2026Lines}. The cNFW model has a constant-density core of radius 1 kpc inside a contracted NFW profile. We extract the curves from the authors' vector figure; numerical ratios obtained by log--log interpolation are approximate graphical comparisons.

The same figure includes the canonical wino and Higgsino line predictions labelled NLO by H.E.S.S. These are reference curves rather than a recalculation for the SU(5) spectra. They incorporate electroweak corrections absent from the fixed-order points~\cite{Baumgart2019,Beneke2022Spectrum}. No theoretical uncertainty band is inferred from the thickness of the published lines. The XENONnT direct-detection curve is taken from its numerical ancillary table, while the LZ and PandaX-4T curves are extracted from their published vector figures.

\section{Results and analysis}\label{sec:results}
\subsection{Sample composition and relic abundance}
The baseline sample is dominated by bino-like points. After imposing the upper relic bound on the quality sample, wino-dominated solutions become predominant. This change reflects the scan and its selections and should not be interpreted as a prior-independent preference of SU(5) for wino dark matter. The sparse coverage of Higgsino and mixed solutions also precludes drawing a reliable boundary for these classes.

\begin{figure}[!htbp]
\centering\includegraphics[width=\textwidth]{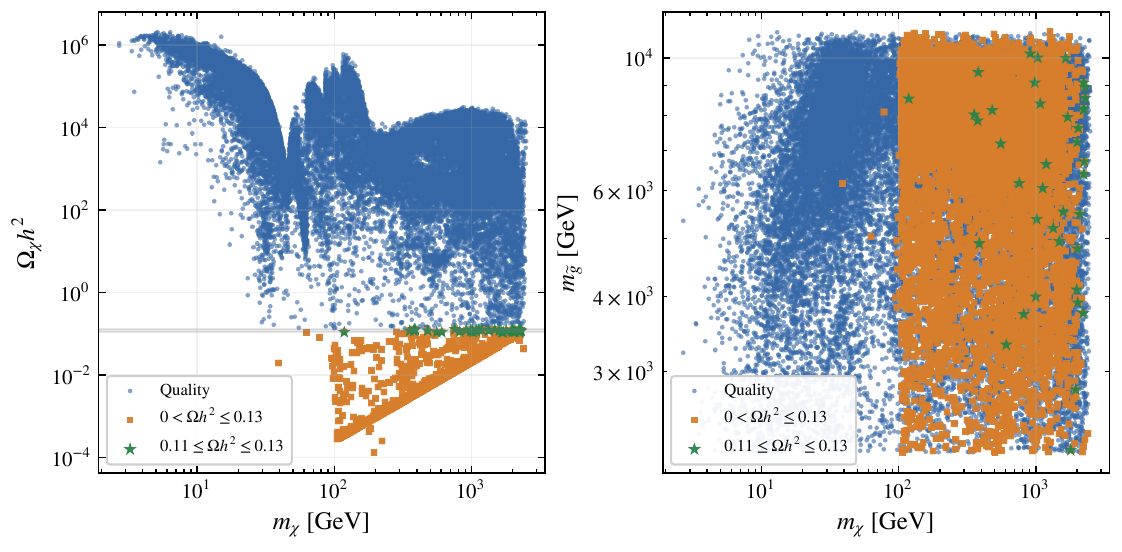}
\caption{Neutralino mass versus thermal relic abundance (left) and gluino mass (right). Blue circles satisfy the quality requirements, orange squares additionally satisfy $0<\om\leq0.13$, and green stars satisfy $0.11\leq\om\leq0.13$. The same nested selections are used throughout.}\label{fig:baseline}
\end{figure}

Figure~\ref{fig:baseline} shows the relic abundance and gluino mass using the same quality and relic selections as the subsequent figures. Overabundant solutions are visible in the quality sample but absent from $\mathcal S_{\rm upper}$. Relatively light neutralinos coexist with multi-TeV gluinos, reflecting the independent electroweak and color gaugino inputs.

\begin{figure}[tb]
\centering\includegraphics[width=\textwidth]{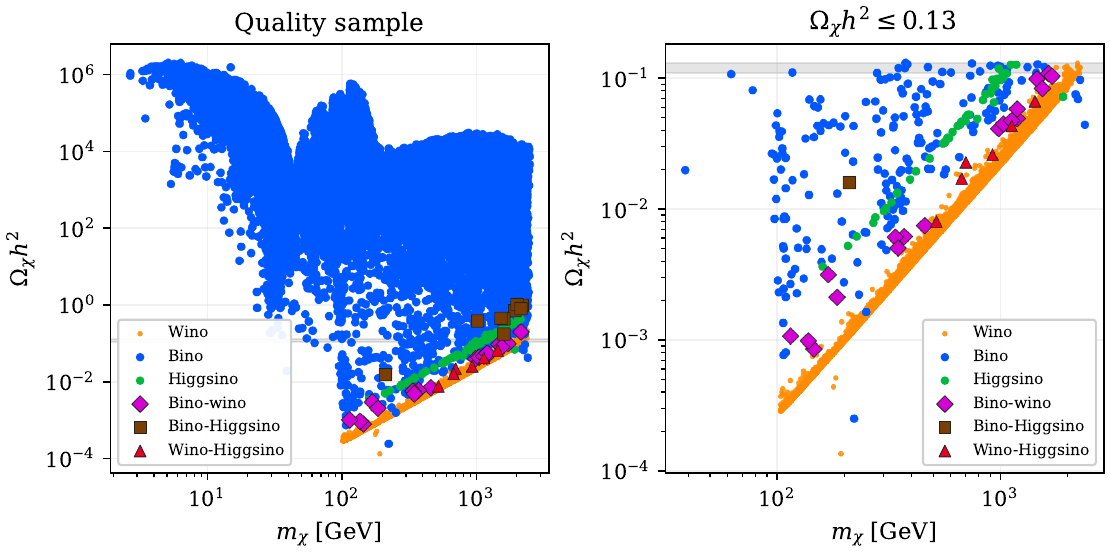}
\caption{Relic abundance by neutralino composition for the quality sample (left) and its $\om\leq0.13$ subset (right). The shaded band denotes the adopted $0.11$--$0.13$ interval. Blue, orange and green circles denote bino-, wino- and Higgsino-dominated solutions, respectively. Magenta diamonds, brown squares and red triangles denote bino--wino, bino--Higgsino and wino--Higgsino mixtures. This composition coding is used throughout the composition-resolved figures.}\label{fig:relic}
\end{figure}

Figure~\ref{fig:relic} shows the quality sample by composition. Most bino-like points overproduce the thermal abundance. Their inefficient annihilation is consistent with the predominantly heavy sfermion spectra and small non-bino admixtures in this sample. A small set of bino-like solutions nevertheless meets the relic requirement. The mass compression of bino benchmark points indicates that nearby electroweakinos can be relevant at freeze-out even when the neutralino wave function is almost entirely bino. This observation also explains why a composition label is not a complete classification of the relic mechanism.

The wino population follows a comparatively regular mass--abundance relation, reaching the saturation interval near $2\TeV$ in the fixed-order thermal calculation. The saturation band also contains nearly pure wino solutions near $1.76$ and $1.92\TeV$, with $f_W=0.990$ and $0.998$, respectively. Their second neutralinos lie only $15.8$ and $64.5\GeV$ above the LSP, so the thermal abundance need not follow the isolated pure-wino relation. A further departure occurs with appreciable bino mixing: a wino-dominated solution at $1.67\TeV$, with $f_W=0.875$ and $f_B=0.125$, reaches the lower edge of the adopted saturation interval. Its composition differs from the canonical pure-wino limit. In the canonical pure-wino limit, Sommerfeld effects shift thermal saturation to approximately $2.9\TeV$~\cite{Beneke2016WinoRelic}. A wino near $2\TeV$ therefore cannot be identified as a Sommerfeld-corrected thermal relic merely because it lies in $\mathcal S_{\rm sat}$ here. Recomputing freeze-out can move such a solution out of the saturation band while retaining it in the upper-bound sample. Near resonances or additional coannihilating states, membership must be evaluated from the corrected effective annihilation rate.

Many solutions have neutralino component fractions close to unity, corresponding to nearly pure bino or wino states. The relic upper-bound sample includes a large population of thermally underabundant winos, while the saturation selection also admits bino and mixed bino--wino solutions. Neutralino composition therefore helps organize the relic-density results but does not by itself determine whether a solution saturates the observed abundance.

Higgsino-dominated solutions occur in both $\mathcal{S}_{\rm upper}$ and $\mathcal{S}_{\rm sat}$. Bino--wino, bino--Higgsino and wino--Higgsino solutions also satisfy the upper relic bound. The mixed bino--wino class extends to $1.69\TeV$ and includes a solution with $f_B=0.319$, $f_W=0.680$ and $\om=0.110$, at the lower edge of the adopted saturation interval. The bino--Higgsino solution has $\mchi=209.4\GeV$, $f_B=0.270$, $f_H=0.717$ and $\om=0.0161$; it is thermally underabundant and falls below the mass range adopted for the H.E.S.S. line comparison. The limited coverage demonstrates that these spectra occur in the stored scan but does not establish the extent or viability of the full mixed-neutralino parameter space. The upper-bound sample includes bino solutions down to $38.7\GeV$ and Higgsino-dominated solutions from approximately $159\GeV$ to $1.89\TeV$. These endpoints describe the sampled spectra rather than mass limits established by all searches. Table~\ref{tab:observables} provides class ranges for the upper-bound sample.

\begin{table}[tb]
\centering\small
\resizebox{\textwidth}{!}{\begin{tabular}{lrrr}
\toprule
Composition & $m_\chi$ range [GeV] & $\Omega_\chi h^2$ range & $\langle\sigma v\rangle_{\rm line}^{\rm FO}$ range \\
\midrule
Bino & $38.7$--$2367$ & $2.56\times10^{-4}$--$0.13$ & $3.61\times10^{-36}$--$9.20\times10^{-29}$ \\
Wino & $103$--$2252$ & $1.36\times10^{-4}$--$0.129$ & $1.58\times10^{-27}$--$9.03\times10^{-27}$ \\
Higgsino & $159$--$1890$ & $3.67\times10^{-3}$--$0.126$ & $1.38\times10^{-29}$--$2.36\times10^{-28}$ \\
Bino--wino & $114$--$1685$ & $8.65\times10^{-4}$--$0.11$ & $1.09\times10^{-28}$--$3.33\times10^{-27}$ \\
Bino--Higgsino & $209$ & $0.0161$ & $6.68\times10^{-29}$ \\
Wino--Higgsino & $513$--$1411$ & $8.13\times10^{-3}$--$0.0674$ & $4.79\times10^{-28}$--$2.54\times10^{-27}$ \\
\bottomrule
\end{tabular}}
\caption{Composition-dependent observables for the quality sample satisfying $\om\leq0.13$. The final column is in $\unitrate$ and uses the convention in Eq.~(\ref{eq:line}). Ranges describe the occupied sample before spectrum-dependent LHC tests; their lower endpoints are not established allowed mass limits. A single value is shown where the endpoints coincide.}\label{tab:observables}
\end{table}

\subsection{Gaugino ratios and compressed spectra}
\begin{figure}[!htbp]
\centering\includegraphics[width=\textwidth]{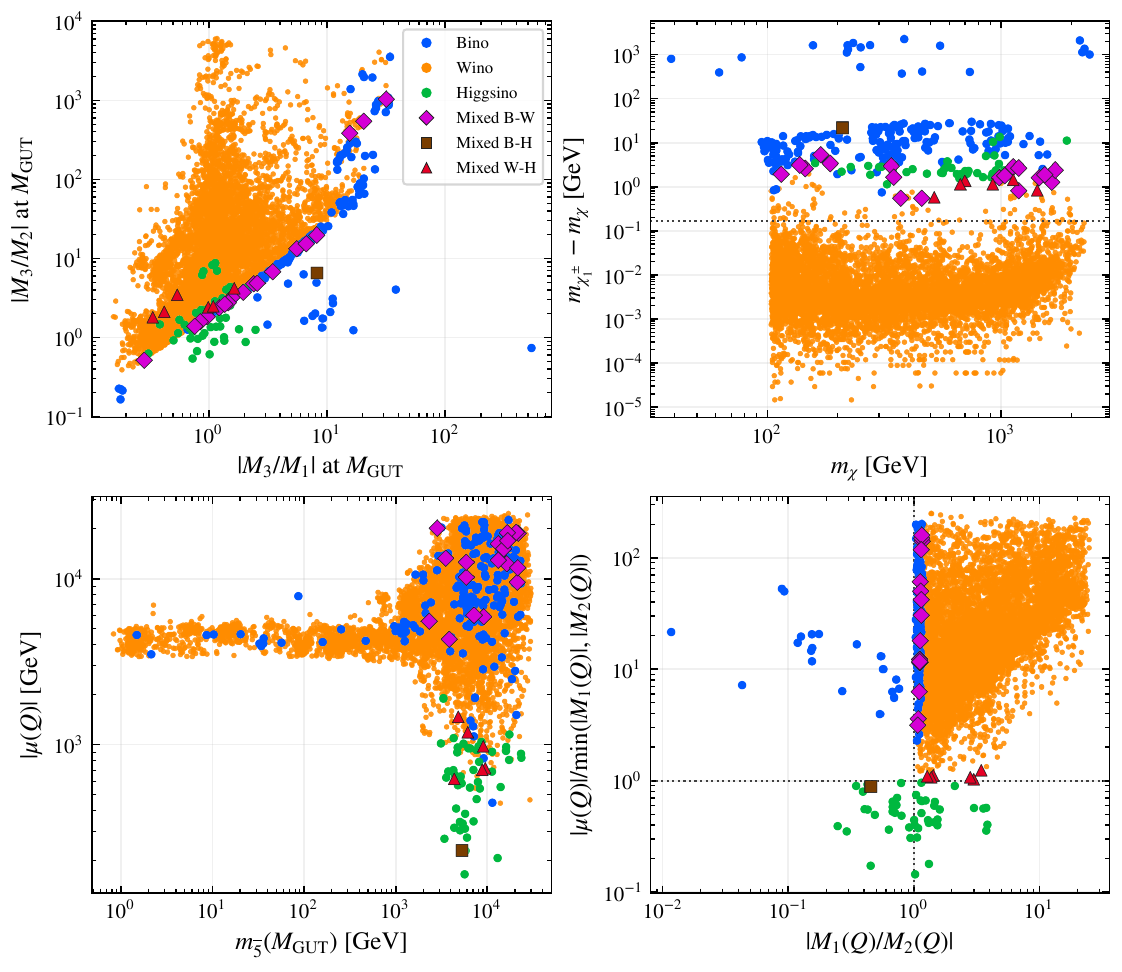}
\caption{Quality upper-bound sample colored by neutralino composition: GUT-scale gaugino ratios (upper left), chargino--neutralino mass splitting (upper right), the common Higgs/scalar boundary mass versus $|\mu(Q)|$ (lower left), and the weak-scale gaugino/Higgsino hierarchy (lower right). Only positive splittings appear on the logarithmic axis. The dotted $0.164\GeV$ line illustrates the heavy pure-wino radiative scale and is not a correction applied to the spectra.}\label{fig:gut}
\end{figure}

Figure~\ref{fig:gut} connects the high-scale inputs to the low-scale electroweakino hierarchy. Its upper-left panel displays $|M_3/M_1|$ and $|M_3/M_2|$ at $\MG$, separating electroweakino composition while retaining a heavy colored sector. SU(5) adds scalar and Higgs-sector relations to this gaugino freedom. In particular, the common Higgs boundary value $m_{\overline5}$ enters the electroweak symmetry-breaking condition~\cite{MartinPrimer},
\[
\mu^2(Q)=\frac{m_{H_d}^2(Q)+\Sigma_d(Q)-[m_{H_u}^2(Q)+\Sigma_u(Q)]\tan^2\beta}{\tan^2\beta-1}-\frac{m_Z^2}{2},
\]
where $\Sigma_{u,d}$ denote effective-potential corrections. Thus $\mu$ is not an independently scanned weak-scale input. Its value depends jointly on the evolved Higgs soft masses, gaugino inputs, trilinear coupling and $\tan\beta$. The lower-left panel shows the sampled $m_{\overline5}$--$|\mu|$ relation; the lower-right panel shows how $|\mu|/\min(|M_1|,|M_2|)$ separates Higgsino from gaugino solutions. The Higgsino-dominated upper-bound solutions have $m_{\overline5}=3.08$--$23.23\TeV$ but $|\mu(Q)|=0.164$--$1.880\TeV$, with $|\mu|/m_{\overline5}=0.016$--$0.574$. They occupy the region in which the evolved Higgs potential yields a Higgsino mass below both electroweak gaugino parameters despite a substantially larger high-scale scalar mass. This is the concrete connection to the common SU(5) Higgs/scalar boundary condition. Heavy scalar inputs alone do not force a large $\mu$; the other high-scale parameters enter the same symmetry-breaking relation. A statistical frequency or a causal effect of $m_{\overline5}$ alone cannot be inferred from this nonuniform scan.

The upper-right panel exhibits the strong compression of the wino-like spectra. The lighter chargino and neutralino share the wino mass parameter, so their near degeneracy is expected qualitatively. Higgsino and mixed points also show electroweakino compression, whereas the bino-like class spans a broader range. For bino benchmarks with a splitting of a few tens of GeV and $\mchi$ near $1\TeV$, the fractional gap is only a few percent. Freeze-out can therefore be sensitive to states other than the lightest neutralino even though the present-day neutralino annihilation signal is small.

\begin{figure}[!htbp]
\centering\includegraphics[width=\textwidth]{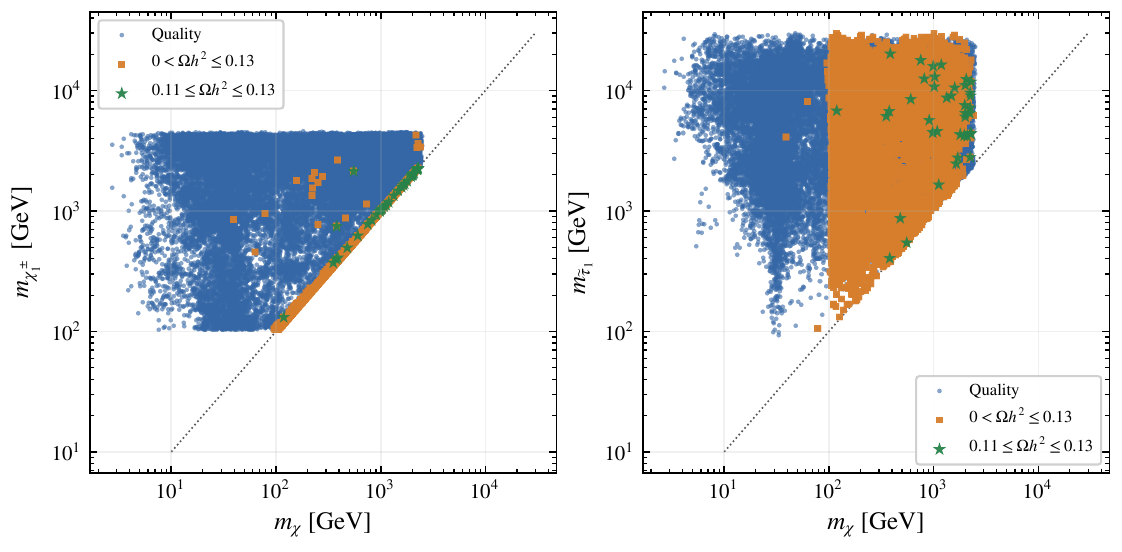}
\caption{Lighter-chargino (left) and lighter-stau (right) masses versus neutralino mass. Markers denote the quality, relic upper-bound and saturation selections as in Fig.~\ref{fig:baseline}. The dotted line denotes mass degeneracy. Proximity to this line is a kinematic coannihilation diagnostic.}\label{fig:masscorrelations}
\end{figure}

Figure~\ref{fig:masscorrelations} shows the chargino and stau mass correlations with the common quality selections. Chargino degeneracy is prominent in $\mathcal S_{\rm upper}$, whereas the stau mass is often much larger. For Point 1, a bino fraction of 0.999 coexists with a chargino only $25.1\GeV$ heavier than the $993\GeV$ LSP. This identifies electroweakino coannihilation as a candidate mechanism. A bino-dominated saturation solution near $545\GeV$ instead has a lighter stau only $5.5\GeV$ above the LSP, identifying stau coannihilation as another candidate mechanism. A quantitative assignment, including any competing resonance or sfermion process, requires the channel contributions to the thermal effective annihilation rate.

The smallest wino splittings are below the approximately $0.16\GeV$ radiative splitting of a heavy pure wino~\cite{Ibe2013WinoSplitting,McKay2018Splitting}. Points 2 and 3 have calculated differences of $0.0056$ and $0.117\GeV$, respectively. These values need a consistent pole-mass calculation before applying disappearing-track searches~\cite{ATLAS2026Tracks,CMS2023Tracks}. A purity threshold alone does not justify replacing a general MSSM splitting by the pure-triplet value: Point 3, for example, has a second neutralino only $17.4\GeV$ above the LSP. For orientation, the ATLAS wino-production interpretation excludes chargino masses up to approximately $880\GeV$ for lifetimes around $1\,\mathrm{ns}$~\cite{ATLAS2026Tracks}. This conditional reach establishes that the low-mass wino population requires collider testing; it does not apply an $880\GeV$ boundary to every wino-like point. A consistent pole spectrum, decay calculation and search interpretation are needed to assign the surviving mass range. Figures~\ref{fig:baseline} and \ref{fig:masscorrelations} therefore display the baseline-selected spectra, including points whose LHC status is unresolved.

\subsection{Direct-detection constraints}
\begin{figure}[!htbp]
\centering\includegraphics[width=\textwidth]{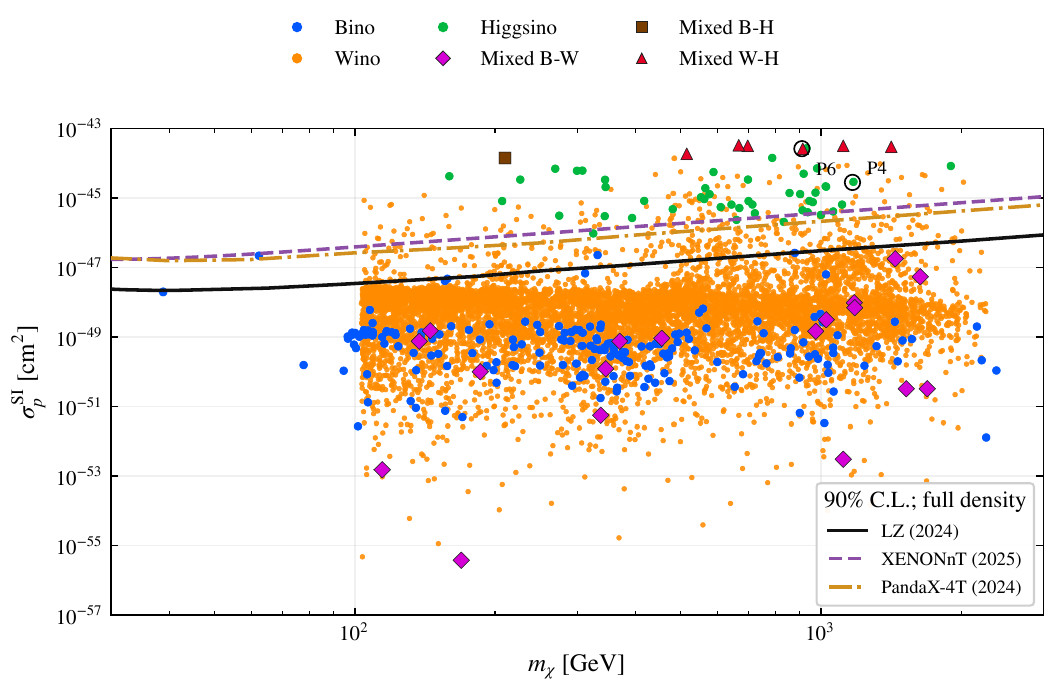}
\caption{Proton spin-independent scattering for $\mathcal S_{\rm upper}$ at full halo density. Lines show observed 90\% C.L. limits from LZ~\cite{LZ2025}, XENONnT~\cite{XENONnT2025} and PandaX-4T~\cite{PandaX2025}; legend years identify the data-release papers. Points above a given experimental curve are excluded by that search under its halo and scattering assumptions; points below it are not excluded by that search. Both populations are retained in the plot. Points 4 and 6 are labelled. The experimental curves use their respective standard halo assumptions. Near $1\TeV$, the XENONnT observed bound is approximately 12 times higher than the LZ bound. These are observed limits, not median sensitivities; the LZ result includes a downward background fluctuation and applies a $-1\sigma$ power constraint~\cite{LZ2025}.}\label{fig:dd}
\end{figure}

Figure~\ref{fig:dd} compares the unmodified proton spin-independent cross sections with the observed xenon limits. At a given neutralino mass, a prediction above an experimental curve is excluded at its stated 90\% C.L., under the full-density and scattering assumptions of the comparison. A prediction below that curve is not excluded by that experiment; it is not thereby established as viable against other searches. The curves are individual experimental bounds, not a combined likelihood, and the plotted sample retains both excluded and nonexcluded predictions.

The composition classes respond differently to these bounds. All sampled Higgsino-dominated solutions lie above LZ, by at least a factor of 7.8, and also above PandaX-4T. XENONnT excludes part of this class but leaves some predictions below its weaker bound; these remain excluded by LZ. The sampled bino--Higgsino and wino--Higgsino solutions are above all three experimental curves. Conversely, the bino--wino solutions lie below all three in the stated scattering calculation. The bino class is largely below the limits. A solution near $62.1\GeV$ exceeds LZ by a factor of 8.4 and PandaX-4T by approximately 1.2, while remaining below XENONnT. A further bino solution near $330\GeV$ exceeds LZ by approximately 2.2 but lies below both XENONnT and PandaX-4T. Thus a bino composition does not by itself ensure compatibility with scattering searches. Wino-dominated predictions occupy both sides of the bounds, so direct detection does not exclude that entire class. These statements describe the sampled spectra and the displayed scattering calculation, not every MSSM realization of a composition label.

This distinction is particularly relevant for Higgsinos. Recent LZ interpretations constrain the allowed gaugino admixture and thus the purity of Higgsino dark matter~\cite{Martin2025Curtain}. Nearly degenerate sleptons can modify the thermal abundance, while relative gaugino signs can suppress scattering through destructive interference~\cite{Yue2025HiggsinoSleptons}. Such spectra illustrate why exclusion of the Higgsino solutions sampled here does not exclude the composition class in general. Recent inelastic Higgsino interpretations of xenon recoil and gamma-ray data instead involve sub-MeV neutralino splittings and transitions to the second neutralino~\cite{Wu2026Higgsino}; they concern a different scattering process from the elastic limits used in Fig.~\ref{fig:dd}.

Points 4 and 6 are excluded under the full-density, approximately isospin-conserving interpretation. Their cross sections are $2.80\times10^{-45}$ and $2.62\times10^{-44}\mathrm{cm^2}$, respectively. At their masses, these exceed the LZ limit by approximately 80 and 950, the XENONnT limit by 6.5 and 78, and the PandaX-4T limit by 11 and 135. Thermal underabundance does not reduce Point 6's scattering rate under Eq.~(\ref{eq:density}).

The mixed bino--Higgsino solution at $209.4\GeV$ has $\sigma_{\rm SI}^p=1.39\times10^{-44}\mathrm{cm^2}$, with proton and neutron cross sections differing by approximately 2\%. Its predicted cross section exceeds the LZ, XENONnT and PandaX-4T limits by approximately $2.1\times10^3$, 174 and 312, respectively. This solution is therefore excluded under the same full-density interpretation despite satisfying the thermal relic upper bound.

The bino benchmark Point 1 lies below these limits, as does the mixed bino--wino Point 5 in the stated scattering calculation. The extremely small cross sections of nearly pure winos require a different interpretation. Electroweak loop contributions survive in the decoupling limit and can dominate a suppressed tree-level amplitude; heavy-WIMP effective theory provides the appropriate reference calculation~\cite{HillSolon2014,ChenHill2020}. Recent one-loop MSSM calculations also find that electroweak corrections and interference can either enhance or suppress wino-like scattering, depending on the spectrum~\cite{Bisal2024WinoSI}. We therefore do not interpret the lowest wino points in Fig.~\ref{fig:dd} as a physical lower bound on the scattering rate. Matching those corrections to each spectrum remains necessary for a precision prediction. In particular, a wino point below a displayed scattering limit is not certified as nonexcluded after loop corrections; points close to a bound also require the appropriate proton and neutron amplitudes in the nuclear response.

\subsection{Line signals under the full-density assumption}
\begin{figure}[!htbp]
\centering\includegraphics[width=\textwidth]{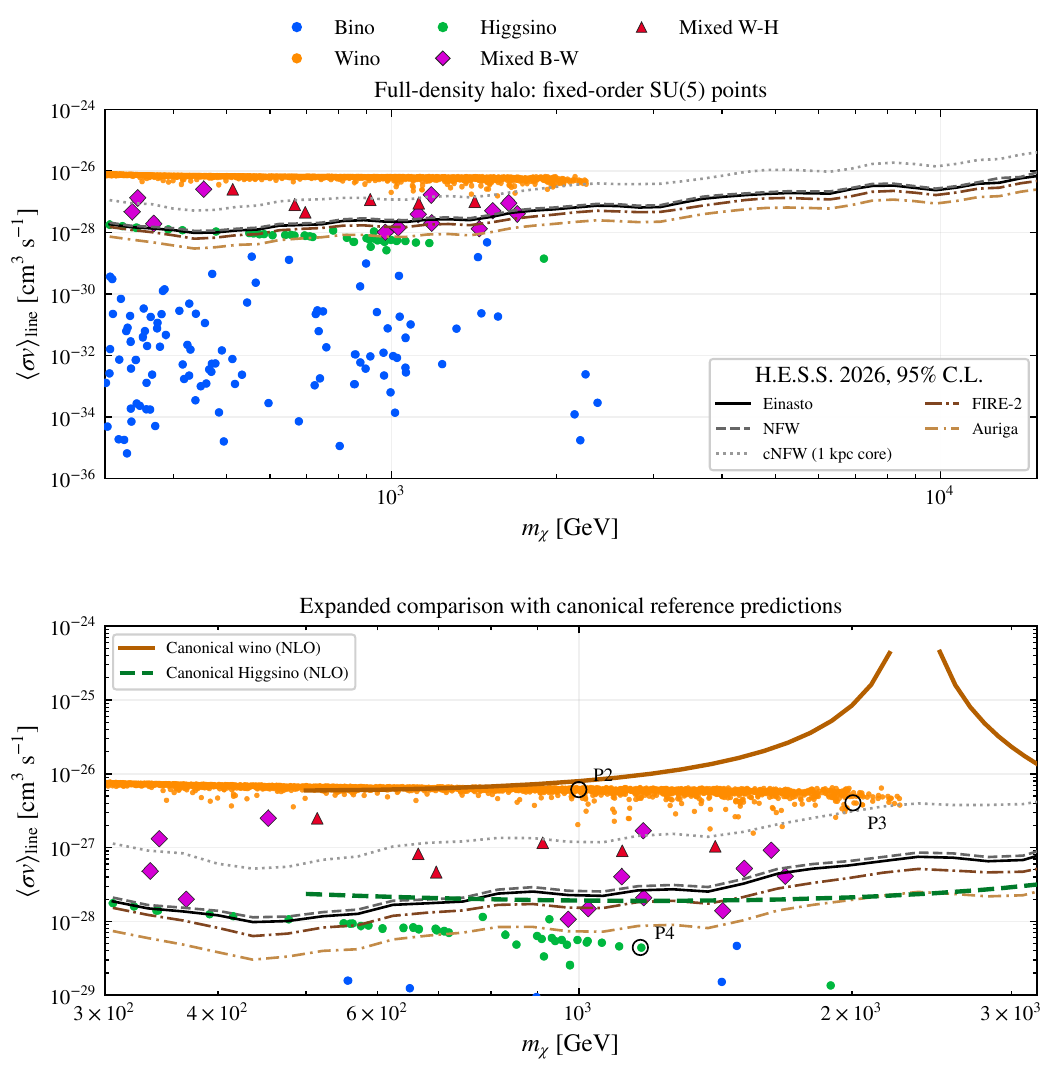}
\caption{Fixed-order SU(5) two-body line rates at full halo density (points) and observed H.E.S.S. 95\% C.L. limits for five halo profiles (lines)~\cite{HESS2026Lines}. The upper panel shows the full signal range. The lower panel expands the comparison and overlays the published canonical wino and Higgsino NLO curves, which are not recalculated for the SU(5) spectra. For a chosen halo profile, predictions above its experimental curve are excluded in the displayed signal approximation; predictions below it are not excluded by that curve. The colored NLO curves are theoretical predictions, not exclusion boundaries. Profiles retain their published normalizations. Curves are extracted from the authors' vector figure; the theoretical curves carry no inferred uncertainty band. The gap near the wino resonance separates the extracted branches of the published reference curve; no interpolation is made across it.}\label{fig:line}
\end{figure}

Figure~\ref{fig:line} compares the fixed-order rates of $\mathcal S_{\rm upper}$ above $300\GeV$ with the H.E.S.S. limits. Both panels assume full neutralino density; the lower panel resolves the experimentally relevant region and includes canonical reference predictions. The two-photon convention in Eq.~(\ref{eq:line}) is used throughout. Its factor of one half is a photon-multiplicity convention, unrelated to relic abundance. For each adopted halo profile, the region above its experimental curve is excluded at 95\% C.L. for the corresponding signal prediction, while the region below is not excluded by that line search. Applied to our points, this is a test of the fixed-order two-body rates; a corrected spectrum-specific exclusion additionally requires the effects discussed below. A point can be excluded for one profile and not excluded for another. We make numerical classifications only within the extracted curves' coverage, which begins near $306\GeV$; lower-mass points are not assigned a H.E.S.S. status by extrapolating these curves.

Across the mass range covered by the extracted limits, all sampled wino-dominated fixed-order rates exceed the Einasto, NFW, FIRE-2 and Auriga bounds. Their smallest Einasto excess is a factor of 4.3. The cNFW comparison is different: the wino-dominated solution near $1.67\TeV$ lies below that curve, at approximately 0.92 times the limit, while remaining above the other four profile bounds. Its $f_W=0.875$ and $f_B=0.125$ distinguish it from a canonical pure wino, and its corrected annihilation signal remains to be calculated. Additional wino-dominated solutions near $1.55$ and $1.91\TeV$ lie at approximately 0.91 times the cNFW limit, with $f_W=0.879$ and $0.971$, respectively, while exceeding the other displayed profile bounds. The remaining wino-dominated rates lie above cNFW, although the closest excess is only about 0.4\%, below the precision warranted by graphical extraction. That near-boundary comparison is not a robust exclusion. This broad fixed-order tension does not by itself supply a corrected exclusion for every spectrum labelled wino dominated.

All sampled bino-dominated rates within the comparison range lie below all five halo-profile limits and are not excluded by this fixed-order line test. The Higgsino rates instead depend more strongly on the adopted profile: all lie below NFW and cNFW, some lie just above Einasto, and portions of the class exceed FIRE-2 and Auriga. The near-Einasto comparisons and missing electroweak corrections preclude a definitive corrected line-search classification from these rates alone. Direct detection already excludes the sampled Higgsino solutions independently. The bino--Higgsino solution near $209\GeV$ is outside this line comparison and is excluded by scattering, not by a H.E.S.S. bound applied here.

The almost pure wino Point 2 has $f_W=0.999914$, $\mchi=999.6\GeV$ and $\sigmav_{\rm line}^{\FO}=6.11\times10^{-27}\unitrate$. This rate is approximately 27 times the extracted Einasto limit and 5.1 times the cNFW limit. Using the rounded published $1\TeV$ Einasto value gives 26.6 instead. The distinction is below the precision warranted by graphical extraction. Full-density canonical winos in this region are excluded; their SU(5) origin does not evade that conclusion. The corrected canonical curve supports this interpretation rather than rescuing the $1\TeV$ wino. At Point 3, the Einasto limit is approximately $5.83\times10^{-28}\unitrate$ at $2005\GeV$, independently recovered from both published H.E.S.S. panels. The observed curve has local structure, so interpolating only between the $1\TeV$ and thermal-wino anchors is not an equivalent check. The fixed-order excess is approximately 6.9 for Einasto and 1.3 for cNFW. Thus Point 3 lies on the excluded side of both curves, as well as the other displayed halo bounds, in the fixed-order comparison. The canonical wino curve rises toward a Sommerfeld resonance near $2.3\TeV$, giving a larger reference signal in this region. Point 3's nearby neutralino and inaccurate chargino splitting can shift the coupled-channel dynamics, so the canonical enhancement cannot be assigned to that spectrum without a corrected calculation. The composition threshold $f_W\geq0.8$ alone is not a canonical-wino exclusion criterion.

Point 4 illustrates the complementary Higgsino issue. Its $4.43\times10^{-29}\unitrate$ fixed-order rate is approximately six times below the Einasto limit at $1169\GeV$. Nevertheless, the canonical NLO Higgsino prediction reaches the experimental sensitivity near its thermal mass~\cite{HESS2026Lines,Beneke2020Higgsino,Beneke2022Spectrum}. A fixed-order point below the curve therefore cannot establish Higgsino viability. Point 4 has a $2.98\GeV$ chargino gap and nonzero gaugino mixing, so the canonical curve is not its corrected prediction. Direct detection already excludes it independently.

The bino Point 1 instead has $\sigmav_{\rm line}^{\FO}=6.43\times10^{-34}\unitrate$, over five orders of magnitude below the Einasto limit at its mass. A small present-day rate is compatible with acceptable freeze-out when nearby unstable sparticles enhance the thermal effective annihilation rate. Point 5 has a larger mixed bino--wino rate of $1.09\times10^{-28}\unitrate$, about one half of the Einasto limit. It is not excluded by the displayed Einasto, NFW, cNFW or FIRE-2 bounds at fixed order, but exceeds the Auriga bound by approximately 1.5. Its line-search status is therefore halo dependent even before a corrected annihilation calculation. Its thermal abundance, $\om=0.0417$, supplies only about one third of the reference cosmological density. Interpreting Point 5 at full halo density therefore requires additional production just as it does for the underabundant wino benchmarks. Falling below the displayed limits establishes neither that production history nor compatibility with other searches. The full-density choice gives stronger detection tests than a subdominant neutralino component; exclusions obtained in this scenario are conditional on that choice.

The mixed classes also illustrate why scattering and line constraints must be read together. All sampled bino--wino solutions within the line-comparison range exceed the Auriga bound, while their status under the other profiles varies. The bino--wino solutions near $454$ and $1175\GeV$ exceed all five line limits at fixed order despite lying below the scattering bounds; Point 5 has the profile-dependent status described above. The mixed bino--wino solution near $1.63\TeV$ also lies below all three scattering limits, but exceeds the Einasto, NFW, FIRE-2 and Auriga line bounds. Its fixed-order rate is approximately 2.2 times the Einasto limit and 0.47 times the cNFW limit. Reaching the adopted thermal saturation band therefore does not remove the halo dependence of its line-search interpretation. The wino--Higgsino solutions exceed Einasto, NFW, FIRE-2 and Auriga. For cNFW, the solution near $513\GeV$ remains above the limit by a factor of 3.9, whereas the other sampled wino--Higgsino solutions lie below it. Point 6 lies at approximately 0.91 times the bound. Point 6 is therefore not excluded by that particular fixed-order line comparison, but is excluded by direct detection regardless of this change in the Galactic Center profile.

Continuum photons provide an additional test of light full-density winos. Fermi-LAT dwarf-spheroidal searches use targets distinct from the Galactic Center and avoid dependence on the Milky Way inner cusp, although dwarf $J$-factor uncertainties remain~\cite{Fermi2024Dwarfs,Cohen2013,Fan2013}. The H.E.S.S. 2022 Inner Galaxy analysis is also a continuum search~\cite{HESS2022Continuum}. More recently, an inner-Galaxy Fermi-LAT analysis has reported exclusion of the canonical thermal wino even for substantially cored halo profiles~\cite{Safdi2025FermiWino}. This result concerns the thermal electroweak multiplet and is not a spectrum-specific reinterpretation of every wino-dominated SU(5) point. Applying continuum constraints to the present spectra requires their annihilation channels and photon yields.

Future CTAO observations could extend the comparison for electroweakino candidates; dedicated wino and Higgsino projections exist~\cite{Rinchiuso2021CTA}. Recent studies examine thermal-Higgsino line signals with CTA and SWGO~\cite{Rodd2024HiggsinoCTA}, including large-zenith-angle Galactic Center observations from CTAO-North~\cite{Abe2025CTAONorth}. Their sensitivity depends on the inner-halo profile and the corrected photon spectrum. These projections motivate further study of Higgsino candidates with sufficiently suppressed scattering, but do not restore the directly excluded Higgsino spectra presented here or imply comparable reach for the very weak bino benchmark.

\subsection{Comparison with a muon magnetic-moment interval}
\begin{figure}[!htbp]
\centering\includegraphics[width=\textwidth]{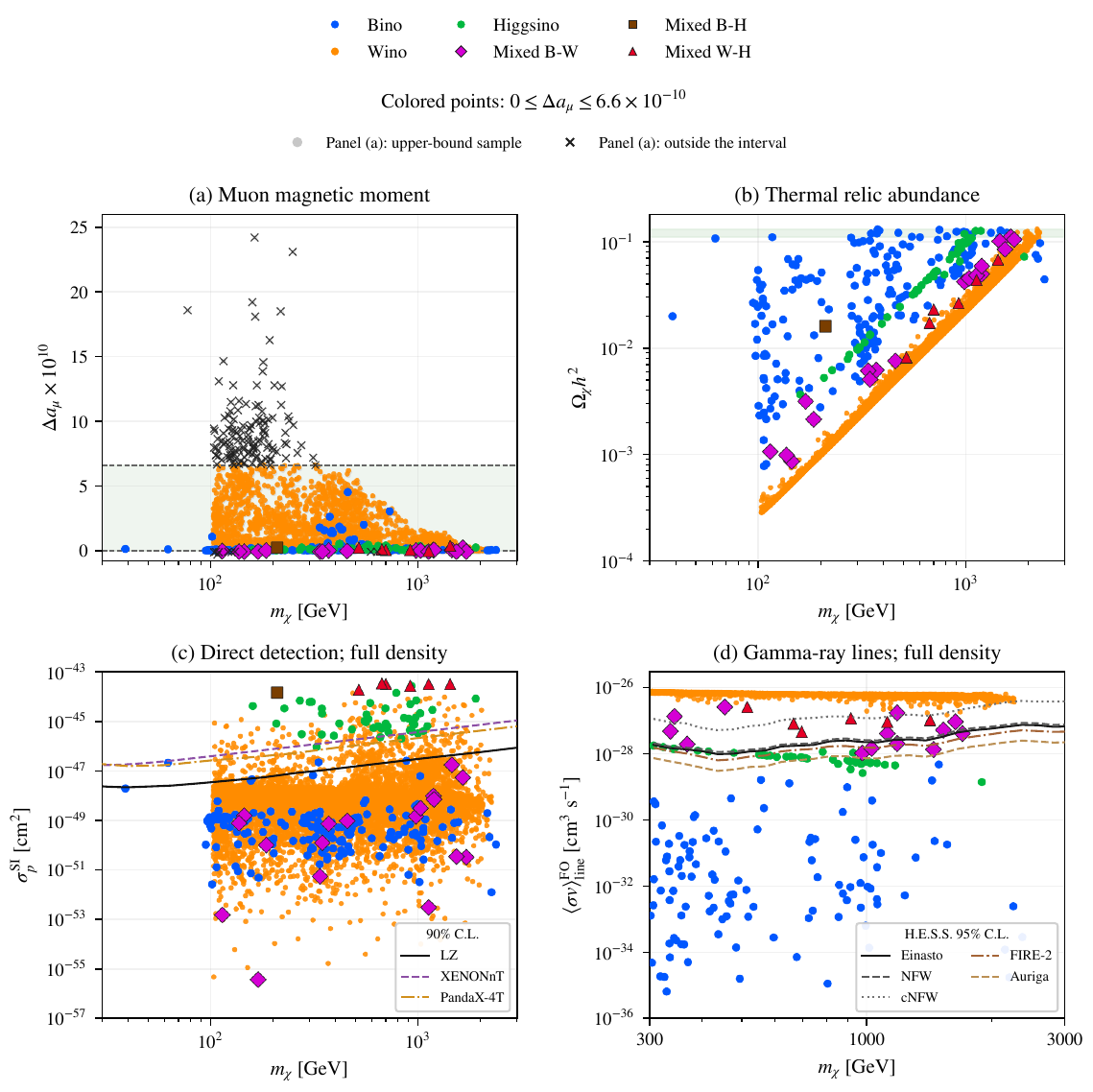}
\caption{Additional muon-$g-2$ comparison for $\mathcal S_{\rm upper}$. Colored points satisfy $0\leq\Delta a_\mu\leq6.6\times10^{-10}$, with colors identifying composition. In panel (a), gray points show the full upper-bound sample and black crosses mark points outside the interval. Panels (b)--(d) contain only points satisfying the muon-$g-2$ interval. Panels show (a) $\Delta a_\mu$, (b) thermal relic density, (c) proton spin-independent scattering and (d) the fixed-order two-photon-equivalent line rate. Shading denotes the chosen $\Delta a_\mu$ interval in (a) and the saturation band in (b). The detection panels retain full neutralino density and the experimental curves of Figs.~\ref{fig:dd} and \ref{fig:line}; panel (d) requires $m_\chi\geq300\GeV$. This additional selection is not applied to the other figures.}\label{fig:gm2subset}
\end{figure}

We examine the subset of $\mathcal S_{\rm upper}$ satisfying
\begin{equation}
0\leq\Delta a_\mu\leq6.6\times10^{-10},\label{eq:gm2subset}
\end{equation}
where $\Delta a_\mu$ denotes the supersymmetric contribution to the muon anomalous magnetic moment. This interval is used only in Fig.~\ref{fig:gm2subset}; the main sample and the preceding figures retain their original selections. Panel (a) compares the selected and unselected solutions through their muon magnetic moment. Panels (b)--(d) show only solutions satisfying Eq.~(\ref{eq:gm2subset}), displaying their thermal relic abundance, spin-independent scattering and fixed-order line rate as functions of the neutralino mass.

The interval retains the full saturation sample and all six benchmarks. The points outside it belong to the bino- and wino-dominated classes, while the Higgsino and mixed-composition solutions in $\mathcal S_{\rm upper}$ remain in the selected subset. The principal detection conclusions therefore persist: the selected Higgsino solutions remain excluded by LZ, and the selected wino-dominated line rates exceed the Einasto bound within the comparison range. The halo-dependent wino solution near $1.67\TeV$ and mixed bino--wino solution near $1.63\TeV$ also remain. Compatibility with Eq.~(\ref{eq:gm2subset}) thus leaves both excluded and nonexcluded predictions under the displayed dark-matter searches.

\subsection{Benchmark points}
Tables~\ref{tab:benchmass} and \ref{tab:benchobs} illustrate distinct compositions and thermal abundances. They include excluded examples to expose the complementarity of line and scattering constraints. Dimensionful parameters and detection observables are kept at the accuracy of the underlying calculation.

\begin{table}[!htbp]
\centering\small
\resizebox{\textwidth}{!}{\setlength{\tabcolsep}{4pt}
\begin{tabular}{lrrrrrr}
\toprule
 & Point 1 & Point 2 & Point 3 & Point 4 & Point 5 & Point 6 \\
\midrule
$m_{10}$ & $11250.3$ & $11406.8$ & $27646.9$ & $16785.4$ & $12772.4$ & $5531.8$ \\
$m_{\overline5}$ & $11108.0$ & $4743.5$ & $12629.8$ & $17208.9$ & $5802.4$ & $8755.6$ \\
$A_0$ & $11125.2$ & $-4143.0$ & $7728.5$ & $-16853.5$ & $-12343.0$ & $9091.8$ \\
$a_0$ & $1.002$ & $-0.8734$ & $0.6119$ & $-0.9793$ & $-2.127$ & $1.038$ \\
$\tan\beta$ & $22.32$ & $32.55$ & $13.88$ & $17.23$ & $16.12$ & $28.64$ \\
$M_1(\MG)$ & $2118.0$ & $4824.9$ & $4189.8$ & $3288.9$ & $1990.7$ & $2640.5$ \\
$M_2(\MG)$ & $1104.0$ & $1086.4$ & $2151.6$ & $1767.7$ & $953.5$ & $989.5$ \\
$M_3(\MG)$ & $-2334.9$ & $-896.6$ & $-3298.0$ & $-2922.0$ & $-4698.8$ & $-4273.0$ \\
\midrule
$\mu(Q)$ & $4108.4$ & $8170.8$ & $18969.1$ & $1143.2$ & $10184.7$ & $973.8$ \\
$m_h$ & $125.1$ & $124.1$ & $125.7$ & $125.5$ & $124.3$ & $126.7$ \\
$m_A$ & $10633.1$ & $7961.5$ & $22386.1$ & $16265.9$ & $11276.1$ & $7014.2$ \\
$m_\chi$ & $993.3$ & $999.6$ & $2005.0$ & $1168.7$ & $971.1$ & $910.4$ \\
$m_{\chi_2^0}$ & $1021.7$ & $2246.5$ & $2022.4$ & $1178.4$ & $980.2$ & $997.6$ \\
$m_{\chi_1^\pm}$ & $1018.3$ & $999.6$ & $2005.1$ & $1171.7$ & $972.7$ & $911.5$ \\
$\Delta m_{\chi^\pm}$ & $25.08$ & $0.005615$ & $0.1168$ & $2.975$ & $1.606$ & $1.15$ \\
$m_{\widetilde g}$ & $5377.3$ & $2306.3$ & $7601.0$ & $6642.4$ & $9828.1$ & $8807.1$ \\
$m_{\widetilde t_1}$ & $6948.4$ & $8509.9$ & $20984.8$ & $10159.1$ & $11512.8$ & $5515.1$ \\
$m_{\widetilde\tau_1}$ & $10651.5$ & $4098.0$ & $12327.3$ & $16224.2$ & $5520.0$ & $4468.0$ \\
$m_{\widetilde\mu_L}$ & $11121.6$ & $4799.3$ & $12579.6$ & $17239.3$ & $5779.4$ & $8792.1$ \\
$m_{\widetilde\mu_R}$ & $11267.9$ & $11543.8$ & $27688.6$ & $16813.2$ & $12786.4$ & $5608.1$ \\
\bottomrule
\end{tabular}}
\caption{High-scale inputs and benchmark spectra. All dimensionful entries are in GeV except where indicated. $m_{H_u}=m_{H_d}=m_{\overline5}$ is understood. The benchmark points are defined in Table~\ref{tab:benchobs}; the tables show the stored calculation and are not certificates of compatibility with all searches.}\label{tab:benchmass}
\end{table}
\begin{table}[!htbp]
\centering\small
\resizebox{\textwidth}{!}{\setlength{\tabcolsep}{4pt}
\begin{tabular}{lrrrrrr}
\toprule
 & Point 1 & Point 2 & Point 3 & Point 4 & Point 5 & Point 6 \\
\midrule
$f_B$ & $0.9989$ & $4.39\!\times\!10^{-6}$ & $0.006781$ & $0.004913$ & $0.7864$ & $0.00231$ \\
$f_W$ & $9.26\!\times\!10^{-4}$ & $0.9999$ & $0.9932$ & $0.01193$ & $0.2135$ & $0.6899$ \\
$f_H$ & $1.25\!\times\!10^{-4}$ & $8.17\!\times\!10^{-5}$ & $1.74\!\times\!10^{-5}$ & $0.9832$ & $4.78\!\times\!10^{-5}$ & $0.3078$ \\
\midrule
$\Omega_\chi h^2$ & $0.122$ & $0.0229$ & $0.117$ & $0.126$ & $0.0417$ & $0.0264$ \\
$\Delta a_\mu$ & $4.95\!\times\!10^{-12}$ & $1.85\!\times\!10^{-11}$ & $1.31\!\times\!10^{-12}$ & $2.11\!\times\!10^{-12}$ & $5.83\!\times\!10^{-12}$ & $1.43\!\times\!10^{-11}$ \\
\midrule
$\sigmav_{\gamma\gamma}^{\FO}$ & $9.16\!\times\!10^{-35}$ & $1.39\!\times\!10^{-27}$ & $9.16\!\times\!10^{-28}$ & $2.49\!\times\!10^{-29}$ & $2.49\!\times\!10^{-29}$ & $3.43\!\times\!10^{-28}$ \\
$\sigmav_{Z\gamma}^{\FO}$ & $1.10\!\times\!10^{-33}$ & $9.44\!\times\!10^{-27}$ & $6.25\!\times\!10^{-27}$ & $3.89\!\times\!10^{-29}$ & $1.69\!\times\!10^{-28}$ & $1.68\!\times\!10^{-27}$ \\
$\sigmav_{\rm line}^{\FO}$ & $6.43\!\times\!10^{-34}$ & $6.11\!\times\!10^{-27}$ & $4.04\!\times\!10^{-27}$ & $4.43\!\times\!10^{-29}$ & $1.09\!\times\!10^{-28}$ & $1.18\!\times\!10^{-27}$ \\
\midrule
$\sigma_{\rm SI}^p$ & $4.59\!\times\!10^{-49}$ & $6.40\!\times\!10^{-50}$ & $7.22\!\times\!10^{-49}$ & $2.80\!\times\!10^{-45}$ & $1.48\!\times\!10^{-49}$ & $2.62\!\times\!10^{-44}$ \\
\bottomrule
\end{tabular}}
\caption{Composition, relic density and detection observables for the points in Table~\ref{tab:benchmass}. Line rates are in $\unitrate$ and proton spin-independent cross sections in $\mathrm{cm^2}$. FO denotes the fixed-order two-body line rate. Points 4 and 6 are excluded by the full-density direct-detection comparison.}\label{tab:benchobs}
\end{table}

Point 1 illustrates bino thermal saturation with a compressed electroweakino spectrum and weak line emission. Points 2 and 3 compare thermally underabundant and nominally saturated winos. Point 4 is a Higgsino saturation example excluded by scattering; Points 5 and 6 have mixed bino--wino and wino--Higgsino composition. Their $Z\gamma/\gamma\gamma$ ratios provide a useful amplitude check: Points 2 and 3 give 6.793 and 6.821, close to the heavy pure-wino relation $2\cot^2\theta_W=6.832$ evaluated with the weak-angle input of the calculation~\cite{Hisano2004}. This agreement tests the relative channel normalization but does not validate missing Sommerfeld corrections.

The remaining candidates require radiatively consistent pole masses and lifetimes, channel-resolved freeze-out with Sommerfeld effects, corrected scattering and photon rates, and spectrum-dependent collider tests. These calculations would determine which points below the displayed detection limits remain viable.

\FloatBarrier
\section{Conclusions}\label{sec:conclusion}
We have investigated neutralino dark matter in SU(5) with nonuniversal gaugino masses, emphasizing the interplay between neutralino composition, thermal relic abundance, and direct- and indirect-detection constraints. Under the full-density interpretation, our main conclusions are:

\begin{enumerate}
\interlinepenalty=10000
\item \textbf{Neutralino composition and spectrum.} Nonuniversal gaugino masses accommodate light electroweakinos alongside a heavy colored sector. The common Higgs and scalar boundary conditions connect the resulting neutralino composition to electroweak symmetry breaking.

\item \textbf{Direct detection.} LZ excludes all Higgsino-dominated solutions in the investigated relic upper-bound sample. The displayed scattering limits also exclude the sampled bino--Higgsino and wino--Higgsino mixtures. Bino and bino--wino solutions include candidates below these limits.

\item \textbf{Gamma-ray lines.} The nearly pure wino benchmark near $1\TeV$ exceeds the H.E.S.S. limits for both Einasto and cored profiles, already at fixed order. For other compositions, the conclusion can depend on the halo profile: the wino-dominated saturation solution near $1.67\TeV$ lies below the cored-profile limit but above the other displayed limits at fixed order.

\item \textbf{Complementarity of detection searches.} Satisfying the relic-density requirement and direct-detection bounds does not ensure compatibility with gamma-ray searches. The bino--wino saturation solution near $1.63\TeV$ illustrates this complementarity. Conversely, the bino saturation benchmark has suppressed scattering and line signals and remains below the displayed experimental limits.

\item \textbf{Muon magnetic moment.} The additional selection $0\leq\Delta a_\mu\leq6.6\times10^{-10}$ retains all saturation-band solutions and all six benchmarks. It therefore leaves the principal detection conclusions unchanged.
\end{enumerate}

These results identify candidates not excluded by the displayed dark-matter searches. Establishing their overall viability requires spectrum-dependent collider tests and refined annihilation calculations; thermally underabundant candidates additionally require a production history consistent with the assumed full dark-matter density.

\FloatBarrier
\section*{Data availability}
The numerical data underlying the figures and tables and the scripts needed to reproduce the analysis are available from the author upon reasonable request.

\begingroup
\small
\setlength{\bibsep}{1pt}
\setlength{\parskip}{0pt}
\interlinepenalty=10000
\bibliographystyle{unsrtnat}
\bibliography{references}
\endgroup

\end{document}